\documentclass[12pt,english,english,english]{article}
\usepackage[T1]{fontenc}
\usepackage[latin1]{inputenc}
\usepackage{babel}

\makeatletter

\providecommand{\LyX}{L\kern-.1667em\lower.25em\hbox{Y}\kern-.125emX\@}

\usepackage[T1]{fontenc}
\usepackage[latin1]{inputenc}
\usepackage{babel}
\usepackage{graphics}

\makeatletter

\usepackage[T1]{fontenc}
\usepackage[latin1]{inputenc}
\usepackage{babel}
\usepackage{graphics}



\makeatother
\begin{document}

\title{Simple Models of Plant Learning and Memory}

\author{Indrani Bose\( ^{*} \) and Rajesh Karmakar\( ^{\#} \)}

\maketitle
\centering Department of Physics\par{}

\centering Bose Institute\par{}

\centering 93/1, A.P.C. Road\par{}

\centering Kolkata-700 009, India\par{}

\hfill{}

\hfill{}

\hfill{}

\hfill{}

\hfill{}

\hfill{}

\raggedright \textbf{Abstract}. Plants are capable of intelligent
responses to complex environmental signals. Learning and memory play
fundamental roles in such responses. Two simple models of plant memory
are proposed based on the calcium-signalling system. The memory states
correspond to steady state distributions of calcium ions.\\
 P.A.C.S. Nos.: 05.10.-a, 05.45.-a, 87.16.-b, 87.17.-d\\
 \hfill{}\\
 \hfill{}\\
 \hfill{}\\
 \hfill{}\\
 \hfill{}\\
 \hfill{}\\
 \hfill{}\\
 \hfill{}\par{}

\raggedright {}{*} \emph{email - indrani@bic.boseinst.ernet.in}\\
 \# \emph{email - rkarmakar2001@yahoo.com}\par{}

\section*{1. Introduction}

Plants have long been regarded as passive organisms since they do
not possess brains, cannot speak and lack in motility. Plants, on
the other hand, constitute 99\% of the biomass of the earth demonstrating
their adaptive ability to survive in widely different environments.
In recent years, there is a growing belief that plants are capable
of intelligent responses to environmental stimuli \cite{key-61,key-63}.
Intelligence can be defined as adaptively variable behaviour in the
lifetime of an organism. It is in this sense that one can speak of
plant intelligence. Plant behaviour is remarkably complex in that
plants exhibit considerable flexibility in their responses, have foresight
and can anticipate future problems\cite{key-64}. To give a specific
example, the parasitic plant dodder can assess the exploitability
of a host soon after their initial contact. The dodder coils about
the host with a specific number of turns and sends in a number of
tentacles depending on its assessment of possible future yield from
the host. Aspects of intelligent behaviour include the ability to
compute, to learn and to retain memory. Similarities between the neuronal
network of brains and calcium signalling systems in plants have been
pointed out \cite{key-63,key-65}. There are many examples of crosstalk,
i.e., connections between the biochemical signalling pathways in plants.
A Boolean representation of the networks of signalling pathways is
possible in terms of well-known logical gates like AND, OR, NAND,
NOR, XOR, and XNOR \cite{key-65}. The Boolean description makes it
possible to draw analogies between plant signalling networks and computing
devices. Recently, an electrical network model of plant intelligence
has been proposed which can perform logical operations \cite{key-66}.

There are close parallels between a neural network and a calcium signalling
system. The former is a network of neurons or nerve cells. The network
has a complex structure as each neuron is connected to a large number
of other neurons. The input to a neuron is in the form of electrical
pulses from the neurons to which it is connected. If the resultant
sum exceeds a threshold value, the neuron fires ({}``on'' state)
and sends out electrical pulses to the other connected neurons through
axons and synaptic junctions \cite{key-67,key-68}. The inactive state
of the neuron is known as the {}``off'' state. In a calcium signalling
system, one has a network of ion channels which may be located on
the outer plasma membrane of the cell or on the membranes of intracellular
vesicles and organelles. Like the neurons, an ion channel can be in
two states, closed or open. The normal concentration of free Ca\( ^{2+} \)
in the cytoplasm is much lower than that in the extra-cellular fluid
and in the intra-cellular vesicles and organelles. When a cell receives
an input signal, a series of biochemical events constituting the signalling
pathway are initiated leading to a rapid rise in the concentration
of the sugar phosphate inosital 1,4,5-triphosphate IP\( _{3} \) in
the cytoplasm. A Ca\( ^{2+} \) ion channel opens when both IP\( _{3} \)
and Ca\( ^{2+} \) bind at appropriate sites of the channel. Through
the open channel, Ca\( ^{2+} \) ions move from the interior to the
exterior of the membrane. The released calcium ions diffuse to neighbouring
channels and open them up giving rise to further calcium release.
The coordinated release of calcium ions gives rise to a calcium wave
in the network. The flow of calcium wave is analogous to the transmission
of electrical pulses in a neural network. Long before the calcium
signalling pathway was identified, J. C. Bose through his pioneering
experiments showed that plant cells are excitable and can transmit
millivolt order electrical signals at the speed of 10 - 40 mm/sec.
\cite{key-69}. Even after Bose's experimental observations, the prevalent
belief amongst plant physiologists was that cell signalling involves
solely chemical diffusion. With the elucidation of the calcium signalling
pathway, it is now well established that the release and subsequent
diffusion of Ca\( ^{2+} \)ions gives rise to propagating electrical
pulses in the cellular network thus vindicating Bose's earlier prediction.

Synaptic transmission is not the exclusive mechanism for neurotransmission
in a neural network. There is increasing evidence that non-synaptic
diffusion neurotransmission plays a significant role in some brain
functions \cite{key-70}. Reaction-diffusion (RD) processes involving
antagonistic neurotransmitters can give rise to spatio-temporal organization
in the neural network. RD systems using only local interactions have
been shown to give rise to wide-ranging phenomena including travelling
waves, oscillations and formation of stationary patterns \cite{key-71,key-72}.
In this paper, we discuss two simple RD-type models of networks of
calcium ion channels and show that spatiotemporal organization takes
place in the steady state. The possible role of such organization
in plant learning and memory is further pointed out. Since learning
and memory are attributes of intelligence, our model studies can be
considered as tentative attempts to obtain insight on the complexities
of plant intelligence.

\section*{2. Reaction - diffusion models of calcium signalling networks}

The first model is a minimal model incorporating some of the essential
features of calcium cell signalling. We consider a network of ion
channels in one dimension (1D). Each channel can exist in two possible
states: inactive and active. In the inactive state a channel is closed
and in the active state a channel is open. An open channel allows
for the release of Ca\( ^{2+} \) ions from the internal store of
the channel. The model dynamics are described in terms of two concentration
variables, \( u_{1} \) and \( u_{2} \), denoting the concentrations
of active channels and calcium ions respectively. Detailed processes,
like IP\( _{3} \) and calcium binding, leading to the opening of
the channel are ignored. An active channel can become inactivated
on the binding of calcium ion to an inactivating site. Ca\( ^{2+} \)
can diffuse through the network of channels and on binding at appropriate
sites can activate/inactivate the channels. The concentration of active
channels at a particular location can change due to changes in the
concentrations of active channels in neighbouring regions. This effect
is represented through a {}``diffusive'' coupling, the diffusion
in this case refers to the propagation of channel activity.

The dynamics of the model are described by the following differential
equations: \begin{equation}
\label{mathed:first-eqn}
\partial _{t}u_{1}=D_{1}\partial _{x}^{2}u_{1}+a_{1}u_{1}-b_{1}u_{2}
\end{equation}

\begin{equation}
\label{mathed:second-eqn}
\partial _{t}u_{2}=D_{2}\partial _{x}^{2}u_{2}+b_{2}u_{1}-a_{2}u_{2}
\end{equation}

\raggedright In Eq.(1), \( D_{1} \) represents the effective {}``diffusion
constant'' for the propagation of channel activity. The rate of change
of the concentration of active channels is assumed to be linearly
proportional to the concentration of active channels. The release
of Ca\( ^{2+} \)ions through open channels can activate further opening
up of calcium channels. This effect is indirectly incorporated through
the second term in Eq. (1). The third term in Eq.(1) represents explicitly
the role of Ca\( ^{2+} \) ions in inactivating ion channels and thereby
decreasing \( \partial _{t}u_{1} \). The first term in Eq.(2) describes
the diffusion of Ca\( ^{2+} \) ions, \( D_{2} \) being the diffusion
constant. The second term shows that \( \partial _{t}u_{2} \) increases
if the concentration \( u_{1} \) of active channels increases. The
third term is a decay term and arises out of the pumping back of Ca\( ^{2+} \)ions
into the internal stores.\par{}

In a remarkable paper in 1952, Turing showed that a RD system involving
a slowly-diffusing activator and a fast-diffusing inhibitor can give
rise to a pattern forming instability \cite{key-73,key-74} in which
stationary patterns are formed in the non-equilibrium steady state.
RD systems with dynamics of the type described in Eqs.(1) and (2)
can exhibit different types of instability including the Turing-type
\cite{key-73}. We provide the details below.

In Eqs.(1) and (2), \( u_{1} \) and \( u_{2} \) represent the concentrations
of the activator (active channels) and inhibitor (Ca\( ^{2+} \))
respectively. The growth of \( u_{1} \) stimulates the growths of
\( u_{1} \) and \( u_{2} \). On the other hand, as the name implies,
a rise in the concentration \( u_{2} \) of the inhibitor inhibits
the growths of \( u_{1} \) and \( u_{2} \). The inhibitor diffuses
further than the activator, i.e., \( D_{2}>D_{1} \). In the absence
of diffusion, the steady state of Eqs.(1) and (2) is given by

\begin{equation}
\label{mathed:third-eqn}
(u_{1s},u_{2s})=(0,0)
\end{equation}

\raggedright The stability conditions from linear stability analysis
are:\par{}

\begin{equation}
\label{mathed:fourth-eqn}
a_{1}-a_{2}<0
\end{equation}

\begin{equation}
\label{mathed:fifth-eqn}
detA=b_{1}b_{2}-a_{1}a_{2}>0
\end{equation}

\raggedright One would now like to determine whether the homogeneous
steady state becomes unstable on inclusion of the diffusion terms
in Eqs.(1) and (2). Define a two-component columns vector \( u \)
with elements \( u_{1} \) and \( u_{2} \). We look for solutions
of the form \( u\sim e^{wt}e^{ikx}. \) Linearising Eqs.(1) and (2)
about the steady state in (3), one gets the following characteristic
equation for \( w: \) \par{}

\begin{equation}
\label{mathed:sixth-eqn}
w^{2}-Tw+\Delta =0
\end{equation}

\raggedright where \par{}

\begin{equation}
\label{mathed:seventh-eqn}
T=a_{1}-a_{2}-(D_{1}+D_{2})k^{2}
\end{equation}

\raggedright and \begin{equation}
\label{mathed:eighth-eqn}
\Delta =b_{1}b_{2}-a_{1}a_{2}-k^{2}(a_{1}D_{2}-a_{2}D_{1})+D_{1}D_{2}k^{4}
\end{equation}
 \par{}

\raggedright The solutions of the characteristic equation are \begin{equation}
\label{mathed:nineth-eqn}
w_{\pm }=\frac{1}{2}[T\pm \sqrt{T^{2}-4\Delta }]
\end{equation}
 \par{}

The steady state \( (u_{1s},u_{2s}) \) is linearly stable if Re \( (w_{\pm } \))
is \( < \) \( 0 \). We have already imposed the conditions, Eqs.(4)
and (5), that the steady state is stable in the absence of diffusion,
i.e., Re \( w_{\pm }(k^{2}=0) \) is \( < \) \( 0 \). If the steady
state is to be unstable to spatial disturbances, Re \( w_{+}(k) \)
\( > \) \( 0 \) for some \( k\neq 0 \). This is true if either
\( T \) is \( > \) \( 0 \) or if \( \Delta (k^{2})<0 \) for some
\( k\neq 0 \). Due to the condition in Eq.(4), \( T \) is always
less than zero. The condition in Eq.(5) demands that the only possibility
for \( \Delta (k^{2}) \) to be negative is if \( (a_{1}D_{2}-a_{2}D_{1})>0. \)
Since \( a_{1}<a_{2} \) (Eq.(4)), one gets the additional condition\begin{equation}
\label{mathed:tenth-eqn}
D_{2}>D_{1}
\end{equation}

\raggedright That is, the inhibitor diffuses faster than the activator.
For \( \Delta (k^{2}) \) to be negative for some non-zero \( k \),
the minimum value of \( \Delta , \) \( \Delta _{min}, \) must be
negative. Differentiating Eq.(8) w.r.t. \( k^{2} \), one gets\begin{equation}
\label{mathed:eleventh-eqn}
\Delta _{min}(k^{2})=b_{1}b_{2}-a_{1}a_{2}-\frac{(a_{1}D_{2}-a_{2}D_{1})^{2}}{4D_{1}D_{2}}
\end{equation}
 \par{}

\raggedright with \begin{equation}
\label{mathed:twelveth-eqn}
k^{2}=k^{2}_{m}=\frac{a_{1}D_{2}-a_{2}D_{1}}{2D_{1}D_{2}}
\end{equation}
 \par{}

\raggedright Thus the condition that \( \Delta (k^{2})<0 \) for some
\( k^{2}\neq 0 \) is \begin{equation}
\label{mathed:thirteenth-eqn}
\frac{(a_{1}D_{2}-a_{2}D_{1})^{2}}{4D_{1}D_{2}}>b_{1}b_{2}-a_{1}a_{2}
\end{equation}
 \par{}

\raggedright To summarize, if the conditions (4), (5), (10) and (13)
are satisfied, the homogeneous steady state is unstable towards a
stationary state with wave number \begin{equation}
\label{mathed:fourteenth-eqn}
k_{m}=[\frac{1}{2}[\frac{a_{1}}{D_{1}}-\frac{a_{2}}{D_{2}}]]^{\frac{1}{2}}
\end{equation}
 \par{}

\raggedright At the bifurcation point, \( \Delta _{min}=0 \). Since
\( \Delta =w_{+}w_{-}, \) this implies that one of the roots of the
characteristic equation is zero. The bifurcation can be brought about
by changing the parameters of the system. In the steady state, stationary
distributions of the activator and inhibitor concentrations are obtained.
\par{}

In the simple network model of calcium ion channels considered by
us, the condition (10) implies that Ca\( ^{2+} \) ions diffuse faster
compared to the {}``diffusion'' of channel activity. This is plausible
in a real calcium signalling system as for a channel to be activated
by neighbouring channels, the released Ca\( ^{2+} \) ions from these
channels have to diffuse to the channel in question followed by the
binding of IP\( _{3} \) and Ca\( ^{2+} \) ions at the appropriate
channel sites. The steady state stationary distributions are analogous
to the attractors (memory states) in the case of a neural network.
Figure 1 shows the steady state distributions of \( u_{1} \) and
\( u_{2} \) in an 1d lattice of 50 sites with periodic boundary conditions.
The distributions have been obtained by solving Eqs.(1) and (2) numerically
on discretizing the derivative terms in the two equations in the Euler
scheme. The values of \( \Delta x \) and \( \Delta t \) have been
chosen to be \( 1 \) and \( 0.01 \) respectively. The discretization
can be treated as a simple approximation to the partial differential
Eqs. (1) and (2). The finite difference equation may alternatively
be treated as a representation of the RD system on a lattice. The
initial state of the lattice is the \( (u_{1},u_{2})=(0,0) \) state
perturbed by small random amounts at all the lattice sites.

In a neural network, learning and memory are interlinked. Networks
learn through reinforcement of pathways connecting signals to response.
One way in which reinforcement can occur is through increasing the
strength of existing synaptic connections between neurons. As pointed
out by Trewavas \cite{key-63}, learning and memory are also interrelated
in the calcium signalling system. On receiving an input signal, the
cellular content of the molecules (like IP\( _{3} \)) participating
in the signalling pathway is increased by a large amount. This is
an example of cellular learning leading to an accelerated information
flux along calcium dependent pathways. Memory of previous signals
in a network can be accessed to transform current signals. A calcium-based
memory corresponds to an inhomogeneous stationary distribution of
Ca\( ^{2+} \) ions. The calcium wave generated by a new signal will
propagate preferentially in those regions where calcium concentration
is high. The final distribution of calcium will be the outcome of
the integration of the current signal modified by a stable, long term
memory. In our model, the inhomogeneous stationary distribution of
calcium which constitutes long term memory is an outcome of the Turing
instability. The signal modified by previous memory activates a unique
combination of effector proteins which ultimately brings about the
desired response to the input . Trewavas has further pointed out that
important aspects of the cell memory are possibly associated with
the cell wall as its removal affects many of the developmental processes
in algae and higher plants. It is well-known that calcium signalling
networks can exhibit limit-cycle behaviour in the form of oscillations
in the Ca\( ^{2+} \) concentration \cite{key-75}. In the model considered
by us, the homogeneous steady state (Eq.(3)) undergoes a Hopf bifurcation
when \( a_{1}>a_{2} \) \cite{key-73}. In this case, \( w_{\pm } \)
(Eq.(9)) are purely imaginary, i.e., \( T=0 \) (Eq.(7)). Since the
solution for \( u \) contains the factor \( e^{wt} \), stable oscillations
are obtained in the system.

Microorganisms like bacteria and bacteriophage share a common feature
with plants, namely, the absence of brains. A recent study \cite{key-76}
has discussed evidence that microorganisms exhibit memory. Memory
implies systems the present state of which is not entirely determined
by present conditions but depends on past history, i.e., on the path
by which the present state is reached. Several examples of history
dependence in biological systems are known \cite{key-76}. Most of
these systems have two or more stable steady states, the so-called
memory states. The choice of a particular state depends on the pathway
followed to reach it. Similar examples of memory in a calcium signalling
system can be given. Ref.\cite{key-73} discusses a simple model of
calcium-stimulated calcium release the kinetics of which is described
by the rate law\begin{equation}
\label{mathed:fifteenth-eqn}
\frac{dx}{dt}=\frac{k_{1}x^{2}}{k_{2}+x^{2}}-k_{3}x
\end{equation}

\raggedright where \( x \) denotes the concentration of Ca\( ^{2+} \)
ions. The first term on the r.h.s describes the autocatalytic release
of calcium ions (more calcium ions imply more open channels leading
to further increases in calcium concentration). The autocatalytic
production saturates for high concentrations of Ca\( ^{2+} \). The
second term represents the pumping back of calcium ions into internal
stores. There are three steady states of the system with concentrations
given by \( x_{1}, \) \( x_{2} \) and \( x_{3} \) respectively.
The first and the third states are stable steady states and the second
state is unstable. For all \( 0<x<x_{2} \), \( x\rightarrow x_{1} \),
whereas for all \( x>x_{2}, \) \( x\rightarrow x_{3}. \) Thus the
signalling system exhibits memory of past history. A neural network
has multiple steady states because transmissions across synaptic junctions
can be both excitatory and inhibitory in nature. A calcium signalling
network can also have multiple steady states since Ca\( ^{2+} \)ions
have both activating and inhibiting effects on the opening of ion
channels.\par{}

\section*{3. Concluding remarks}

The models of calcium signalling networks studied in this paper are
toy models meant to illustrate the origin of memory states in plants
in analogy with similar states in neural networks. In reality, the
ion channels have a complex three subunit structure and a channel
is open only if all the three subunits are simultaneously open \cite{key-78}.
Furthermore, the dynamics are governed by non-linear rate equations.
Work on a more realistic model of the calcium signalling network is
in progress and the results will be reported elsewhere. Recently some
discrete stochastic models of calcium dynamics have been proposed
\cite{key-75,key-79}. These models address other interesting aspects
of calcium signalling networks. In one such model defined on a 1d
lattice, a set of probabilities for the opening/closing of calcium
channels is assumed to depend on the calcium concentration. By increasing
the number of channels/site, a transition from a non-propagating region
of activity to a propagating one occurs. The transition belongs to
the directed percolation class of similar transitions. To sum up,
calcium signalling networks present us with a rich array of problems
ranging from cellular learning/memory to novel phenomena arising out
of activated dynamics.

\centering ACKNOWLEDGEMENT\par{}

\raggedright I. B thanks Bikas K. Chakrabarti, Bharati Ghosh, Dibyendu
Sen Gupta and P. C. Sen for useful discussions. R. K. is supported
by the Council of Scientific and Industrial Research, India under
Sanction No. 9/15 (239) / 2002 - EMR - 1.\par{}

\end{document}